\documentclass[12pt]{article}
\usepackage{import}
\usepackage{custom_header}

\begin{document}

\title{One-step Targeted Maximum Likelihood Estimation for Time-to-event Outcomes}
\author{Weixin Cai and Mark J. van der Laan \\
 \it Division of Biostatistics, University of California, Berkeley, CA, USA}
\date{\today}

\maketitle

\begin{abstract}
Researchers in observational survival analysis are interested in not only estimating survival curve nonparametrically but also having statistical inference for the parameter. We consider right-censored failure time data where we observe $n$ independent and identically distributed observations of a vector random variable consisting of baseline covariates, a binary treatment at baseline, a survival time subject to right censoring, and the censoring indicator. We assume the baseline covariates are allowed to affect the treatment and censoring so that an estimator that ignores covariate information would be inconsistent. The goal is to use these data to estimate the counterfactual average survival curve of the population if all subjects are assigned the same treatment at baseline. Existing observational survival analysis methods do not result in monotone survival curve estimators, which inflates their variance. In this paper, we present a one-step Targeted Maximum Likelihood Estimator (TMLE) for estimating the counterfactual average survival curve. We show that this new TMLE can be executed via recursion in small local updates. We demonstrate the finite sample performance of this one-step TMLE in simulations and an application to a monoclonal gammopathy data.

\end{abstract}

\section{Introduction}

Researchers in observational survival analysis are interested in not only
estimating survival curve nonparametrically but also having statistical
inference for the survival curve as a whole. We consider right-censored failure
time data where we observe $n$ independent and identically distributed
observations of a vector random variable consisting of baseline covariates, a
binary treatment at baseline, a survival time subject to right censoring, and
the censoring indicator. We assume the baseline covariates are allowed to affect
the treatment and censoring so that an estimator that ignores covariate
information would be inconsistent. The goal is to use these data to estimate the
counterfactual average survival curve of the population if all subjects are
assigned the same treatment at baseline.

Existing methods such as inverse probability of censoring weighted (IPCW)
estimator, estimating equations (EE) and targeted maximum likelihood estimator
(TMLE) do not produce a monotone estimator of the curve, which translates to
large variance. The reason is that these estimators separately estimate the
survival curve for each time point. The IPCW estimator
\citep{robins1992recovery} re-weights the observed data by the inverse of the
product of the propensity score and censoring probability before applying a
standard estimation method. The EE estimator \citep{hubbard2000nonparametric} is
a locally efficient and double robust estimator, which improves the IPCW by
adding the sample mean of the efficient influence curve. EE is more efficient
than IPCW when the conditional distribution of failure given treatment and
baseline covariates is consistently estimated \citep{hubbard2000nonparametric}.
For IPCW, its consistency relies on correctly estimating the conditional
survival function of censoring. In contrast, EE is doubly robust in the sense
that if either the conditional failure distribution or both propensity score and
conditional censoring probability is correctly estimated, then the EE estimator
will be consistent \citep{hubbard2000nonparametric}. TMLE is a plug-in doubly
robust and locally efficient estimator and is shown to be better than the IPCW
and EE methods \citep{Moore_vdL_2009, Stitelman_2010}. In contrast to these
methods, TMLE performs an adjustment on the estimate of the data distribution
prior to applying the parameter mapping thus always respecting the parameter
space (probabilities falling inside [0,1]) \citep[Chapter 6 of ][]{tmlebook1}. As a result, TMLE
is a plug-in estimator that is more robust in finite samples than EE. While TMLE
works well to improve the statistical efficiency of EE, it can still give rise
to a non-monotone survival curve. The reason is that both EE and TMLE are built
on efficiency theory for univariate parameters. As a result, their solutions for
estimating the survival curve is a collection of univariate survival probability
estimators.

In this article, we propose a TMLE that targets the survival curve as a whole,
while still preserving the performance of the point-wise TMLE for the survival
curve at a point. Due to the joint targeting, the resulting estimator is a
monotone function. The method we propose is built upon the recent advancement of
TMLE theory called one-step TMLE \citep{vdl2016onestep}. This powerful
framework estimates the entire survival curve and ensures monotonicity. We also
discover that the proposed new algorithm is more stable and computationally more
efficient than classic TMLE. We also give a new insight into one-step TMLE by comparing it to the high-dimensional penalized regression literature, which will shed light on the superior finite sample performance of our method.

\paragraph*{Organization of paper}
We start in Section \ref{sec:notaion} by defining the right-censored data, stating the parameter
of interest, and reviewing the efficient influence curve of the parameter. In
Section \ref{sec:initial_estimate} we review nonparametric regressions used in observational survival
analysis, and in Section \ref{sec:review_ipcw} we formally review the IPCW, EE, and classic TMLE
estimators. In Section \ref{sub:tmle} we present intuition on why EE and classic TMLE do
not always produce a monotonically decreasing survival curve. We use this
intuition to build a TMLE that ensures monotonicity in Section \ref{sec:one_step}. In Section \ref{sec:simulation}
we present a simulation study demonstrating the finite sample performance of the
estimators, and in Section \ref{sec:data_analysis} we present an applied example.

\section{Statistical formulation of estimation of the survival curve} \label{sec:notaion}

Let the full data be $X_i = (W_i, A_i, C_{1i}, C_{0i}, T_{1i}, T_{0i}), i = 1,
..., n$, where $W$ is a vector of baseline covariates, $A \in \{0, 1\}$ is
binary treatment assigned at baseline, $T_1$ is the failure time under
treatment, $T_0$ is the failure time under control, $C_1$ is the censoring time
under treatment, $C_0$ is the censoring time under control. Our observed data is
$O_i = (W_i, A_i, \Delta_i, \widetilde T_{A_i}) \sim^{i.i.d} P_0 \in
\mathcal{M}$ for $i = 1, ..., n$, where $\widetilde T \triangleq \min(T_A, C_A)$
is the last measurement time of the subject, and $\Delta \triangleq I(T_A
\leqslant C_A)$ is the censoring indicator. $P_0$ denotes the true probability
distribution of $O$, and we use $p_0$ to denote the true probability density.
$\mathcal{M}$ is the model space of distributions which is believed to be
nonparametric.

The causal parameter is the marginal survival curve in the whole population
where every subject is under the same treatment $$Pr(T_a > t), t = 1, ...,
t_{max},$$ where $T_a$ is the counterfactual failure time one would have
observed had an individual's treatment been set, possibly contrary to fact, to
treatment level $a$. The parameter can be causally identified from the observed
data under the assumptions: (a) no unmeasured confounder, (b) coarsening at
random (the joint variable of censoring and treatment is conditionally
independent of the full data given the observed data), and (c) positivity assumption
\citep{hubbard2000nonparametric,gill1997car,prop_score1983}. After causal identification, our task is reduced to estimating the
statistical parameter $$\Psi_{A=a}(P)(t) = E[Pr(T > t | A=a, W)], t = 1, ...,
t_{max}.$$ This $\Psi: \mathcal{M} \to [0,1]^{t_{max}}$ is a mapping from model
space $\mathcal{M}$ to the parameter space of survival probabilities. $\Psi(P)$
is whole survival curve and $\Psi(P)(t)$ is the survival probability at $t$. For
the rest of the paper, we demonstrate estimators focusing on example in this
parameter family, the treatment-specific marginal survival curve $\Psi_{A=1}$.
Symmetric arguments can be made about $\Psi_{A=0}$, and thus all transformations
of the two parameters (such as difference of two counterfactual survival
probabilities). The components needed to plug into $\Psi \equiv \Psi_{A=1}$ for
the estimand are the conditional survival curve for failure event and the
distribution of $W$, which need to be learned from the observed data. For
performing observational survival analysis, the conditional survival function
for censoring and propensity score also need to be estimated.
Under the causal identification assumptions, the probability density under $P$ factorizes as follows:
\begin{align}
\begin{split}
p(O) = q_W(W)g(W)\prod_{t \leqslant \widetilde T}
\lambda_N(t|A,W)^{dN(t)} (1 - \lambda_N(t|A,W))^{1 - dN(t)} \\
\prod_{t \leqslant \widetilde T} \lambda_{A_c}(t|A,W)^{dA_c(t)} (1 - \lambda
_{A_c}(t|A,W))^{1 - dA_c(t)},
\label{eq:likelihood}
\end{split}
\end{align}
where ${q_W}$ is the density of probability distribution of $W$; $g(W) = P(A|W)$ is the propensity score; $\lambda_N(t|A,W)$ and $\lambda_{A_c}(t|A,W)$ are the conditional hazards of the failure event and censoring event; $dN(t)$ and $dA_c(t)$ are the counting process indicators of the failure event and censoring event. We will formally define them in Section \ref{sec:initial_estimate}.

\subsection{Efficient influence curve}

The EE and TMLE methods to be discussed in this paper are built around the
parameter's efficient influence curve offer a straightforward approach to
estimation. Bickel et al. (1993) show that a regular estimator for a statistical
parameter in a semiparametric model is asymptotically efficient (i.e., the
estimator has minimal asymptotic variance), if it is asymptotically linear with
influence curve (influence curve) equal to the efficient influence curve
(EIF). Under our model space $\mathcal{M}$, \cite{hubbard2000nonparametric} derived the EIF
for $\Psi$ as

\begin{align}
\begin{split}
\label{eq:D1}
D^*_t(P) & = \sum_{k \leqslant t} {h_t(g_{0,A},S_{0,A_c},S_{0,N})}
(k,A,W)\biggl[ I(\widetilde T = k,\Delta  = 1) - \\
& I(\widetilde T \geqslant k)\lambda_{0, N}(k|A = 1,W) \biggr] + S_{0,N}(t|A
= 1,W) - \Psi_d( P )(t) \\
& \equiv D^*_{1,t}(g_{0,A},S_{0,A_c},S_{0,N}) + D^*_{2,t}(P),
\end{split}
\end{align}
where
\begin{align}
\label{eq:clever}
h_t(g_{0,A},S_{0,A_c},S_{0,N})(k,A,W) =  - \frac{I(A = 1) I(k
\leqslant t)}{g_{0,A}(A = 1|W)S_{0,A_c}(k\_|A,W )} \frac{S_{0,N}( t|A,W
)}{S_{0,N}(k|A,W )}.
\end{align}
\section{Nonparametric estimation of components for observational survival analysis methods} \label{sec:initial_estimate}

After causal identification, existing observational survival analysis methods
depend on estimating four components nonparametrically: (1) conditional
survival function for failure event given treatment and confounders, (2)
conditional survival function for censoring event given treatment and
confounders, (3) propensity score of treatment given confounders, and (4)
distribution of confounders in the population of interest.

\paragraph{Conditional survival function for failure event}\mbox{}\\
The conditional survival function is estimated by first estimating the
conditional hazard of the failure event, and then transforming into the
conditional survival function. The definition of the conditional hazard is
\begin{align}
\lambda_N(t|A,W) &= P(\tilde{T} = t,\Delta = 1 | \tilde{T} \geqslant t, A, W) \\
& = P( {dN(t) = 1|N(t-1)=0, A_c(t-1)=0,A,W} ),
\label{eq:hazard_failure}
\end{align}
where $N(t) = I(\widetilde T \leqslant t,\Delta  = 1)$, $A_c(t) = I(\widetilde T \leqslant t,\Delta  = 0)$ and
\begin{align}
    dN(t) =
    \begin{cases}
        1, \text{if }N(t) = 1\text{ and }N(t-1) = 0\\
        0, \text{otherwise},
    \end{cases}
\end{align}
\begin{align}
    dA_c(t) =
    \begin{cases}
        1, \text{if }A_c(t) = 1\text{ and }A_c(t-1) = 0\\
        0, \text{otherwise}.
    \end{cases}
\end{align}
The definition (\ref{eq:hazard_failure}) gives guidance of how to construct a
classification task and estimate the conditional hazard. We first construct a
training data where each subject $O_i$ is mapped into $t_{max}$ rows in a new
data with covariates $(dN(t)_i, N(t-1)_i, A_c(t-1)_i, A_i, W_i, t), t =
1,...,t_{max}$. Estimating the conditional hazard now becomes classification of
$dN(t)_i$, using $(N(t-1)_i, A_c(t-1)_i, A_i, W_i, t)$ as features, performed on
the subset of rows that satisfy the criteria $N(t-1)_i = 0$ and $A_c(t-1)_i =
0$. Note that we include an extra feature $t$ into the design matrix and pool
data from all $t = 1, ..., t_{max}$ into one classification model. Empirically
we found that smoothing over $t$ accelerates the training of classification
algorithms. We follow the common standard to transform the conditional hazard
into the conditional survival function:
\begin{align*}
S_N(t |A,W) = P(T > t | A, W) = \prod_{k = 1}^{t} {(1 - \lambda_N (k|A,W))}.
\end{align*}

\paragraph*{Conditional survival function for censoring event}\mbox{}\\
The conditional survival function for censoring is estimated in the same fashion
as that for the failure event, while swapping the role of $N$ and $A_c$ when
constructing the classification dataset.
\begin{align*}
\lambda_{A_c}(t|A,W) &= P(\tilde{T} = t,\Delta = 0 | \tilde{T} \geqslant t, A, W) \\
& = P( {dA_c(t) = 1|N(t-1)=0, A_c(t-1)=0,A,W} ),\\
S_{A_c}(t |A,W) &= P(C > t | A, W) = \prod_{k = 1}^{t} {(1 - \lambda_{A_c} (k|A,W))}.
\end{align*}

\paragraph*{Propensity score}\mbox{}\\
We estimate the propensity score by running a classification of $A$ against $W$
as features.
\begin{align*}
g(W) = P(A = 1| W).
\end{align*}

\paragraph*{Distribution of confounders}\mbox{}\\
We model the joint distribution of confounders using the empirical probability distribution
of $W_1, ..., W_n$, which we denote as $Q_{n,W}$.
\section{Review of existing observational survival analysis methods} \label{sec:review_ipcw}

\subsection{Inverse probability of censoring weighted estimator}

The inverse probability of censoring weighted (IPCW) estimator re-weights the
observed data by the inverse of the product of the propensity score and
censoring probability in order to make the treatment arms among the uncensored
subjects comparable with respect to confounders, and then applies standard estimation as
if treatment was randomized and censoring was non-informative. The IPCW
estimator for $\psi_0(t)$ is
\begin{align}
    \psi_{n,IPCW}(t) = \frac{1}{n} \sum_{i = 1}^n \frac{I(\tilde T_i > t, \Delta_i = 1, A_i = 1)}{S_{A_c}(\tilde T_i | W_i, A = 1) g(W_i)}.
\end{align}

\subsection{Estimating equations method}

The estimating equation (EE) method is an asymptotically linear estimator based
on solving the efficient influence curve equation: 
\begin{align}
\frac{1}{n} \sum_{i = 1}^n D^*_t(P_n)(O_i) = 0.
\label{eq:eif_equation}
\end{align}
We remind readers that a regular estimator $\psi_n$ of $\psi_0$ is
asymptotically linear if and only if $\psi_n - \psi_0$ behave approximately as
an empirical mean of a mean-zero, finite-variance function of the observed $O$,
where $\psi_0 = \Psi(P_0), \psi_n = \Psi(P_n)$ are the estimand and the estimate.
This function is referred to as the estimator's influence curve
(\ref{eq:D1}). The EE method is one way for constructing estimators with
user-specified influence curve, which applies an EIF-based correction to the
plug-in estimate. Once the empirical influence curve is evaluated for each
observation, the EE method is the IPCW estimator added to the sample mean of EIF
evaluated on each observation.
\begin{align}
    \psi_{n,EE}(t) = \psi_{n,IPCW}(t) + \frac{1}{n} \sum_{i = 1}^n D^*_{t,n}(O_i),
\end{align}
where $D^*_{t,n}(O_i) = D^*_{t}(P_n)(O_i) = D^*_{t}(g_n, Q_n)(O_i)$ is calculated by plugging in the initial estimators of $Q_n = (Q_{n,W}, S_{n,N})$ and $g_n = (g_{n,A}, S_{n,C})$ into $D^*_{t}$ and evaluate at $O_i$.

\subsection{Targeted maximum likelihood estimator} \label{sub:tmle}

TMLE is a general framework for constructing plug-in estimators that satisfy
user-specified equations, which in our case is the EIF equation
(\ref{eq:eif_equation}). It is a plug-in estimator in the sense that the
estimators for $S_{N}(t|A=1, W)$ can be plugged into the mapping $\Psi$ to
calculate an estimate as $$\Psi(Q_n)(t) = \frac{1}{n}
\sum_{i = 1}^{n} S_{n,N}(t|A=1, W_i).$$ Since TMLE updates parts of the
likelihood before applying the parameter mapping, it is guaranteed to fall
inside the range $[0,1]$ of the survival probability.

For the TMLE of $\Psi(t)$, the method is implemented in two steps. First,
initial estimators of the four components are generated by user in Section
\ref{sec:initial_estimate}. Subsequently, the initial estimators are carefully
modified such that (i) the modified estimators inherit desirable properties of
the initial estimators (e.g., their rate of convergence); and (ii) relevant,
user-specified equations are satisfied. For the present problem, the conditional
survival function of failure event is iteratively updated to form a targeted
estimator $\Psi^*_n = \Psi(P^*_n) = \Psi(g_{n},S_{n,A_c},S^*_{n,N})$, such that
the EIF estimating equation $\frac{1}{n} \sum_{i = 1}^n D^*_t(P^*_n)(O_i) = 0$
is satisfied. This can be achieved, for example, by defining a logistic
regression working model for the failure event conditional hazard, with
$\text{logit}(\lambda_{(k)}) = \text{logit}(\lambda_{n,N}(k| A = 1, W))$ as an offset, no
intercept term, and a single covariate $h_{(k)}$, regressed onto the binary
outcome $N_{(k)} = I(\widetilde T = k,\Delta  = 1)$. For each $(k,W), k = 1,
..., t_{max}$, we define this covariate as $h_{(k)} =
h_t(g_{n,A},S_{n,A_c},S_{n,N})(k, 1, W)$. The maximum likelihood estimator
$\varepsilon_n$ of the regression coefficient $\varepsilon$ associated with the
covariate $h_{(k)}$ is estimated (via iterative re-weighted least squares).  For
each $W$, we define the so-called targeted $S^*_{n,N}$ as the conditional
survival function transformed from the targeted conditional hazard
$\lambda^*_{n,N}(k| A=1, W) = \text{expit}\{\text{logit}(\lambda_{(k)}) + \varepsilon_n
h_{(k)}\}$. For notation simplicity, we use $P_n$ and $P_n^*$ for the initial
and targeted distribution of $P_0$, where $P_n = (g_n, S_{n,A_c}, Q_{n,W},
\lambda_{n,N})$ and $P_n^* = (g_n, S_{n,A_c}, Q_{n,W}, \lambda_{n,N}^*)$.The
$g_n$, $S_{n,A_c}$ and $Q_{n,W}$ are never updated because they are tangent to
our statistical parameter of interest and only $\lambda_{n,N}$ is updated. Here
we illustrate one iteration of the targeting step and assume it has converged,
while in practice one iteration is not enough and one might have to iterate many
times until $\|\varepsilon_n\|$ is small or explicitly check the value of
$\frac{1}{n} \sum_{i = 1}^{n} D^*_t(P^*_n)(O_i)$ smaller than a threshold. It is
straightforward to show that the score of the coefficient $\varepsilon$ at
$\varepsilon = 0$ evaluated at a typical observation $O$, equals
$D^*_t(P_n)(O)$; thus, we may deduce that the EIF estimating equation is
satisfied by the updated failure event conditional survival function
$S^*_{n,N}$. The TMLE $\Psi^*_n$ of the treatment-specific marginal survival
curve is computed as the plug-in estimator based on the modified conditional
survival function, $\Psi(Q^*_n)(t) = \Psi(S^*_{n,N}, Q_{n,W})(t) =
\int{S^*_{n,N}(u| A=1, W)dQ_{n,W}(u)} = \frac{1}{n} \sum_{i = 1}^{n}
S^*_{n,N}(t| A=1, W_i)$.

Under regularity conditions on the initial estimates $S_{n,N}$, $S_{n,A_c}$ and
$g_n$, the TMLE is regular and asymptotically linear \cite{tmlebook1}, so $\sqrt{n} (\Psi^*_n(t) - \Psi_0(t)) \to^d N(0, \sigma^2)$. When
$S_{n,N}$, $S_{n,A_c}$ and $g_n$ are consistent estimators for $S_{0,N}$,
$S_{0,A_c}$ and $g_0$, the variance $\sigma^2$ is the variance of the EIF. In
order to estimate the variance $\sigma^2$, we can use an estimate of the sample
variance of the EIF. Wald type hypothesis tests can be performed, and confidence
intervals can be constructed with the estimated variance $\sigma^2_n$. TMLE is
also double robust in the sense that the TMLE is consistent if either (a) the
propensity score $g(W)$ and the censoring event conditional survival probability
$S_{A_c}(A,W)$ are consistently estimated or (b) the failure event conditional
survival probability $S_{N}(A,W)$ is consistently estimated.

\paragraph{Motivation: Why existing TMLE for survival curve is not monotone}\mbox{}\\
The existing TMLE for the marginal treatment-specific survival curve can be
viewed as an application of TMLE in Section \ref{sub:tmle} repeated for survival
probabilities at $t = 1, ..., t_{max}$. The steps for the TMLE algorithm
outlined in Section \ref{sub:tmle} can be summarized in the following
pseudo-code:

\begin{algorithm}[H]
\caption{classic TMLE for survival curve}
\KwData{initial estimator: conditional hazard for failure event, conditional survival curve for censoring event, propensity score}
\KwResult{TMLE for the counter-factual marginal survival curve $\Psi_{A=1}$}
\For{$t = 1, ..., t_{max}$}{
    initialize $S^{(0)} = S_{n,N}$ with the initial estimator for the survival curve of the failure event\;
    $j = 0$\;
    \While{True}{
        \For{$i = 1, ..., n$}{
            \For{$k = 1,...,\tilde T_i$}{
                evaluate $h^{(j)}_{(i,k)} = h_t(g_{n,A},S_{n,A_c},S^{(j)})(k, A_i, W_i)$\;
                evaluate $N_{(i,k)} = I(\widetilde T_i = k,\Delta  = 1)$\;
                evaluate $\lambda^{(j)}_{(i,k)} = \lambda^{(j)}(k,A = 1,W_i)$\;
            }
        }
        concatenate into vectors $h^{(j)}$, $N$ and $\lambda^{(j)}$\;
        get $\hat \varepsilon$ by running a logistic regression $\text{logit} N = \text{logit}(\lambda^{(j)}) + \varepsilon h^{(j)}$\;
        evaluate $\lambda^{(j+1)} = \text{expit}(\text{logit}(\lambda^{(j)}) + \hat\varepsilon h^{(j)})$\;
        transform to $S^{(j+1)}$\;
        $j += 1$\;
        \If{$|\hat \varepsilon| \le 1e-3$}{\textbf{break}}
    }
    $\Psi^*(t) = \frac{1}{n} \sum_{i = 1}^n S^{(j)}_i(t)$\;
}
concatenate the $\Psi^*(t)$ to get the entire curve $\Psi^*(t), t = 1,...,t_{max}$\;
\end{algorithm}

Note that the method creates $t_{max}$ different $\lambda^*_{n,N,\tilde{t}}, \tilde{t} = 1, ..., t_{max}$ for each $\Psi(\tilde{t})$ task, therefore transforming the multiple $\lambda^*_{n,N,\tilde{t}}$ into survival probabilities does not create a monotone decreasing survival curve.

\section{One-step TMLE targeting the entire survival curve} \label{sec:one_step}

The logistic submodel we use in the previous section is also called the local least favorable submodel (LLFM) around $\lambda_{n,N}$:
\begin{align}
    \text{logit}(\lambda_{n,N, \varepsilon}(k| A=1, W)) = \text{logit}(\lambda_{n,N}(k| A = 1, W)) + \varepsilon h_{(k)},
    \label{llfm}
\end{align}
because it has the property that
\begin{align*}
\frac{d}{d\varepsilon} \log\frac{dP_{n,\varepsilon}}{dP} |_{\varepsilon = 0} = D^*_t(P_n),
\end{align*}
where $D^*_t(P_n)$ is the short notation for the EIF at
$(g_{n,A},S_{n,A_c},\lambda_{n,N})$ and $P_{n,\varepsilon}$ is the distribution
at $(g_{n,A},S_{n,A_c},\lambda_{n,N,\varepsilon})$. This is a key result that
ensures TMLE is solving the EIF estimating equation by running a logistic
regression along the submodel (\ref{llfm}), but it also implies that the results
hold only if we use the submodel around $\varepsilon = 0$, that is, we don't
update along the submodel with a large step size $\varepsilon_n$. Doing a
logistic regression on this submodel (\ref{llfm}), however, does not guarantee
that $\varepsilon_n \approx 0$. This intuition explains why doing TMLE on a
high-dimensional parameter can often lead to diverging results, because TMLE is
an iterative algorithm and because the first few iterations usually involve
large step sizes.

\cite{vdl2016onestep} proposed a novel targeting step to modify the initial
estimators called one-step TMLE. The idea is that since the gradient equals the
EIF only locally when we update the initial estimators, one-step TMLE only
performs the update locally. If we make the step size small enough, the submodel has
the property that at any $\varepsilon$
\begin{align*}
\frac{d}{d\varepsilon} \log\frac{dP_{n,\varepsilon}}{dP} = D^*_t(P_{n,\varepsilon}) = D^*(\lambda_{n,N, \varepsilon}, Q_{n,W}, g_n).
\end{align*}
This submodel is known as the universal least favorable submodel (ULFM) around $\lambda_{n,N}$, which takes the form
\begin{align}
    \text{logit}(\lambda_{n,N, \varepsilon}(k| A=1, W))= \text{logit}(\lambda_{n,N}(k| A=1, W)) + \int_0^\varepsilon h_t(g_{n,A},S_{n,A_c},S_{n,N,x})(k, 1, W)dx.
    \label{ulfm}
\end{align}
This theoretical formulation gives an insight into how this methodology works,
but is not useful when analyze our survival curve problem because it involves
integration of a complex function of $S_{n,N,x}$ (which itself is a function of
$\lambda_{n,N,x}$).

In execution, the one-step TMLE is carried out by many LLFMs (performed in
logistic regressions) with small step sizes. The one-step TMLE updates in small
steps locally along LLFM, making sure only using the update direction ($h_t(.)$)
that is optimal around the current probability density. One-step TMLE also
allows the analyst to update the conditional hazard for all points on the
survival curve (or any high-dimensional parameter in general), so that the
conditional hazard can be transformed into a monotone survival curve after the
algorithm. To do this, one replaces the univariate $h_t(.)(k,1,W)$ in
(\ref{llfm}) with a high dimensional vector $\vec{h}_t(.) = (h_t(.)(1,1,W), ...,
h_t(.)(t_{max},1,W))$, each one corresponding to the clever covariate of
survival probability at one time point. Fitting the high-dimensional logistic
regression will not hurt the performance since we never update with large step
size. Another way to view the one-step TMLE is that the logistic regression we
used within classic TMLE is replaced with a logistic ridge regression, where
the coefficient L-2 norm is constrained to be smaller than a tiny value.
Because the logistic ridge regression generally outperforms classic logistic
regression in high dimensions, the one-step TMLE is better than classic TMLE for
high-dimensional target parameters. Given the same input and output, one-step
TMLE leads to a new targeting procedure. The essential steps becomes the
following pseudo-code, where the differences between one-step TMLE and classic
TMLE are highlighted.

\begin{algorithm}[H]
\caption{one-step TMLE for the survival curve}
\KwData{initial estimator: conditional hazard for failure event, conditional survival curve for censoring event, propensity score}
\KwResult{TMLE for the counter-factual marginal survival curve $\Psi_{A=1}$}
initialize $S^{(0)} = S_{n,N}$ with the initial estimator for the survival curve of the failure event\;
$j = 0$\;
\While{True}{
    \For{$i = 1, ..., n$}{
        \For{$k = 1,...,t_{max}$}{
            evaluate $N_{(i,k)} = I(\widetilde T_i = k,\Delta  = 1)$\;
            evaluate $\lambda^{(j)}_{(i,k)} = \lambda^{(j)}(k,A = 1,W_i)$\;
            \HiLi \For{$t' = 1, ..., t_{max}$}{
                \HiLi evaluate $h^{(j)}_{(i, k, t')} = h_{t'}(g_{n,A},S_{n,A_c},S^{(j)})(k, A_i, W_i)$\;
            }
            \HiLi concatenate into vector $\vec{h}^{(j)}_{(i,k)}$\;
        }
    }
    \HiLi concatenate along $(i,k)$ indices (by row) into vectors $N$, $\lambda^{(j)}$ and matrix $\vec{h}^{(j)}$\;
    \HiLi get $\hat \vec{\varepsilon}$ by running a logistic ridge regression $\text{logit} N = \text{logit}(\lambda^{(j)}) + \vec{\varepsilon} \vec{h}^{(j)}$ subject to $\| \vec{\varepsilon} \| \leq 1e-2$\;
    evaluate $\lambda^{(j+1)} = \text{expit}(\text{logit}(\lambda^{(j)}) + \hat\vec{\varepsilon} \vec{h}^{(j)})$\;
    transform to $S^{(j+1)}$\;
    $j += 1$\;
    \If{$\|\hat \vec \varepsilon\| \le 1e-3$}{\textbf{break}}
}
\HiLi $\Psi^*(t) = \frac{1}{n} \sum_{i = 1}^n S^{(j)}_i(t), t = 1,...,t_{max}$\;
\end{algorithm}
Note: With abuse of notation, we define $h_{(i, k, t')} = h_{t'}(g_{n,A},S_{n,A_c},S_{n,N})(k, A_i, W_i)$ to include an additional subscript $t'$ referring to the clever covariate for estimating $\Psi(t')$ evaluated at observation $O_i$.

\paragraph*{Inference}\mbox{}\\
The statistical inference of iterative and one-step TMLE at a single time point
can be done in the same procedure. The TMLE estimators, both iterative and
one-step, solve the efficient influence curve equation:
\begin{align*}
\frac{1}{n} \sum_{i = 1}^n D^*_t(P^*_n)(O_i) = 0, t = 1, ..., t_{max}.
\end{align*}
Thus, if all components are consistent and under regularity conditions, TMLE is
asymptotically linear with influence curve $D^*_t(P_0)$
\citep{vdl2003unified}. Based on this result, TMLE inference is based on the empirical
variance of the efficient influence curve $D^*_t(P^*_n)$, assuming the initial estimators $(S_N, g_A, S_{A_c})$ are consistent. Thus, the
asymptotic variance of $n^{1/2}(\psi_n^*(t) - \psi_0(t))$ is estimated by:
\begin{align*}
\hat\sigma_t^2 = \frac{1}{n} \sum\limits_{i = 1}^n {D^*_t}^2(P^*_n)(O_i).
\end{align*}
Now a valid $100 \times (1 - \alpha )\%$ confidence interval
is constructed under the normal distribution in the following way: $$\psi_n^*(t) \pm q_{1 - \alpha /2} \frac{\hat \sigma_t}{\sqrt n},$$
where $q_\beta$ is the $\beta$-quantile of the standard normal distribution.

\paragraph*{Simultaneous confidence interval}\mbox{}\\
The simultaneous confidence bands for the survival curve estimates can be
similarly constructed based on asymptotic linearity of the TMLE uniform in all
time points considered. Inference for $\vec{\psi}^*_n$, the vector of survival
probabilities at $t_{max}$ time points, a vector parameter, is also based on the
empirical variance of the efficient influence curve $\vec{D}^*$ itself at the
limit of $(S_N^*,g_A,S_{A_c})$. The asymptotic variance of
$n^{1/2}(\vec{\psi}^*_n - \vec{\psi}_0)$ may be consistently estimated by the
$t_{max}$ by $t_{max}$ empirical covariance matrix of the efficient influence
curve:
\begin{align*}
\hat \Sigma = \frac{1}{n}\sum\limits_{i = 1}^n \vec{D}^*(P^*_n)(O_i) \{\vec{D}^*(P^*_n)(O_i)\}^ \top .
\end{align*}
By multivariate central limit theorem, we have
\begin{align}
n^{1/2} (\vec{\psi}^*_n - \vec{\psi}_0)\mathop  \to \limits^d N(0, \Sigma_0).
\end{align}
 As a result, an approximate $100
\times (1 - \alpha )\% $ simultaneous confidence band is constructed such that
for each $\psi(t)$, the $t^{th}$ component of $\vec{\psi}$, the region
is given by
\begin{align*}
\psi^*_n(t) \pm q_{1 - \alpha} \hat\Sigma^{1/2}(t)/\sqrt n,
\end{align*}
where $\hat \Sigma ^{1/2}(t)$ is the $(t,t)$-th entry in the empirical
covariance matrix, thus the empirical variance of $D_t^*$. $q_{1 - \alpha}$ is
an estimate of the $1 - \alpha$ quantile of $\max_t \sqrt{n} | \psi^*_n(t) -
\psi_0(t) |/\widehat \Sigma ^{1/2}(t)$. Here we need to use that the latter
random variable behaves as the max over $t$ of $Z(t)$, where $Z\sim N(0,\rho )$
follows $t_{max}$-dimensional gaussian and $\rho$ is the correlation matrix of
the vector influence curve $\vec{D}^*(P^*_n)(O_i)$. We simulate Monte-Carlo
samples of $Z$ and calculate $q_{1 - \alpha }$ using the empirical $1-\alpha$
quantile of ${\max _t}| Z |$ of the random samples. Due to actual weak
convergence of the standardized TMLE as a random function in function space
endowed with supremum norm, these simultaneous confidence bands are valid even
as we take a finer and finer grid of time points as $n$ increases.

\section{Simulation}
\label{sec:simulation}

\begin{figure}[!ht]
\centering
\includegraphics[width=.9\textwidth]{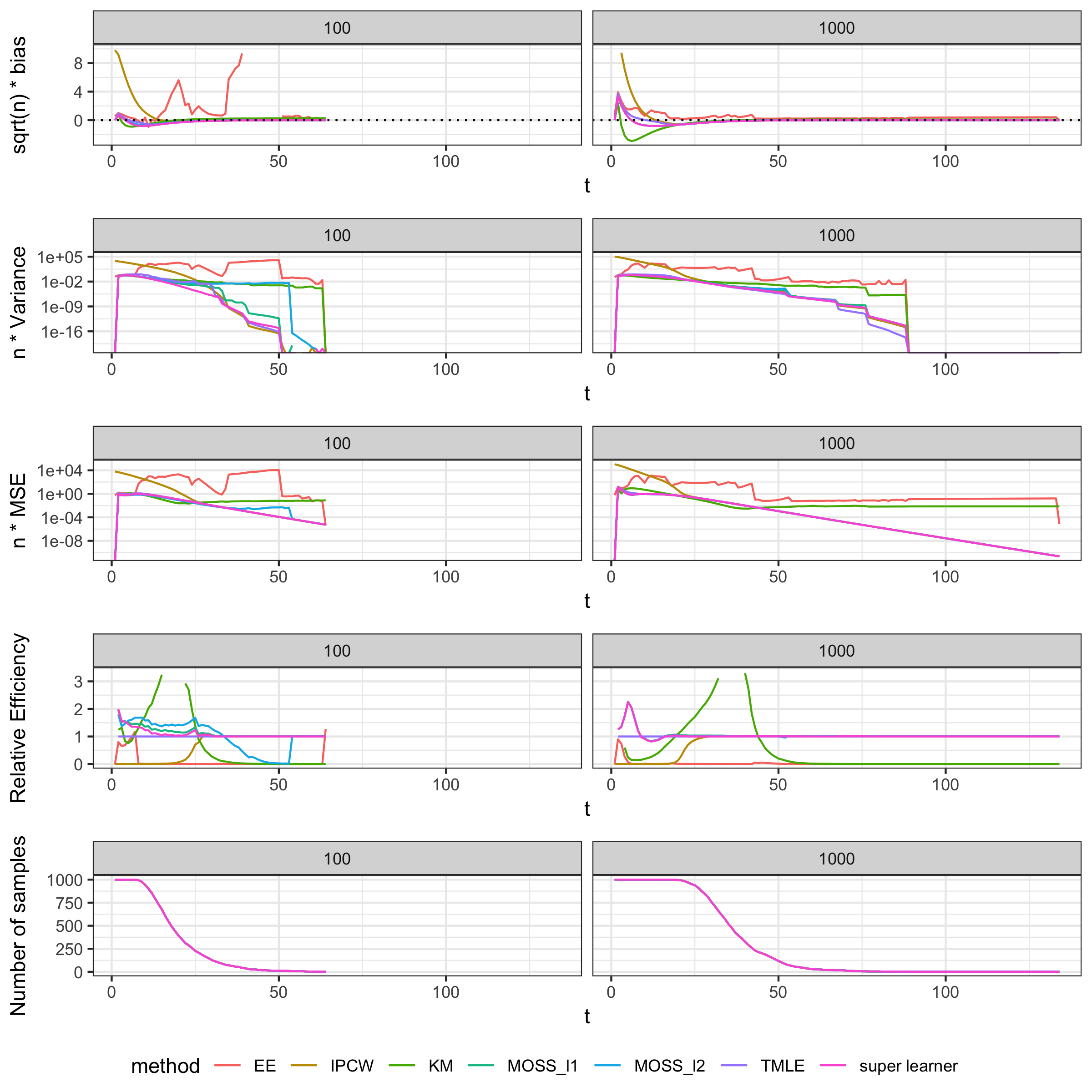}
\caption{Results for comparing different survival curve estimators at all time points. Row 1 is bias, row 2 is variance, row 3 is MSE, row 4 is relative efficiency (larger than 1 means more efficient than iterative TMLE), row 5 is  the number of simulations where follow up time is at least t. Within each row, the left plot is under sample size 100 and the right plot is under sample size 1000. Note the relative efficiency value larger than 5 are truncated so that the plot range around [0,1] can be easily interpreted}
\label{fig:sim}
\end{figure}

To provide an example of the finite sample properties of the estimators discussed in Sections \ref{sec:review_ipcw} and \ref{sec:one_step}, we simulate a univariate continuous baseline covariate $W$, a binary exposure $A$, a survival outcome $T$ with censoring time $C$. We simulate data from the following data-generating distribution so that $T$, $A$, and $C$ are confounded by $W$:
\begin{align*}
W &\sim \text{Unif}(0, 1.5),\\
A &\sim \text{Bernoulli}( {0.4 + 0.5 I\{W > 0.75\} } ),\\
T &\sim \text{log-normal} (\mu = 2 - W + A, \sigma = 0.01),\\
C &\sim \text{Weibull}( 1 + 0.5W, 75 ).
\end{align*}
To analyze the above simulated data, we estimate the survival curves under the treatment and control groups. 
For sample sizes n = 100 and 1000, we simulated 1000 Monte-Carlo repetitions from the previous DGD, and estimated $\Psi_{A=1}(P_0)$ and $\Psi_{A=0}(P_0)$ using the following estimators: Kaplan-Meier; plug-in SuperLearner estimator of the conditional survival curve \citep{superlearner}; IPCW; EE; classic (iterative) TMLE; one-step TMLE targeting the whole curve.
As initial estimators of the components of the likelihood $(g_0, S_{0,A_c}, \lambda_{0,N})$, we used SuperLearner classification combining multiple classification algorithms so that we know the estimates will be consistent. The SuperLearner library includes generalized linear model \citep{glm}, generalized additive model \citep{gam}, and multivariate adaptive regression splines \citep{earth}. We used empirical distribution $Q_{n,W}$ to estimate $Q_{0,W}$.
One-step TML estimation was performed using the R function `MOSS\_hazard` in the open-source package MOSS \citep{MOSS}, and the code that reproduces this simulation is presented in Web Appendix.
The average and variance of the estimates across the 1000 samples was computed as an approximation to the expectation and variance of the estimator, respectively. We report the bias, variance, mean-squared error (MSE) of different estimators in Figure \ref{fig:sim}, and we use the MSEs to further calculate the relative efficiencies (RE) against iterative TMLE for all estimators:
\begin{align*}
RE_{\text{estimator}}(t) = \frac{MSE_{\text{iterative TMLE}}(t)}{MSE_{\text{estimator}} (t)}, t = 1,...,t_{max}.
\end{align*}

\begin{figure}[!ht]
\centering
(a)\includegraphics[width=.4\textwidth]{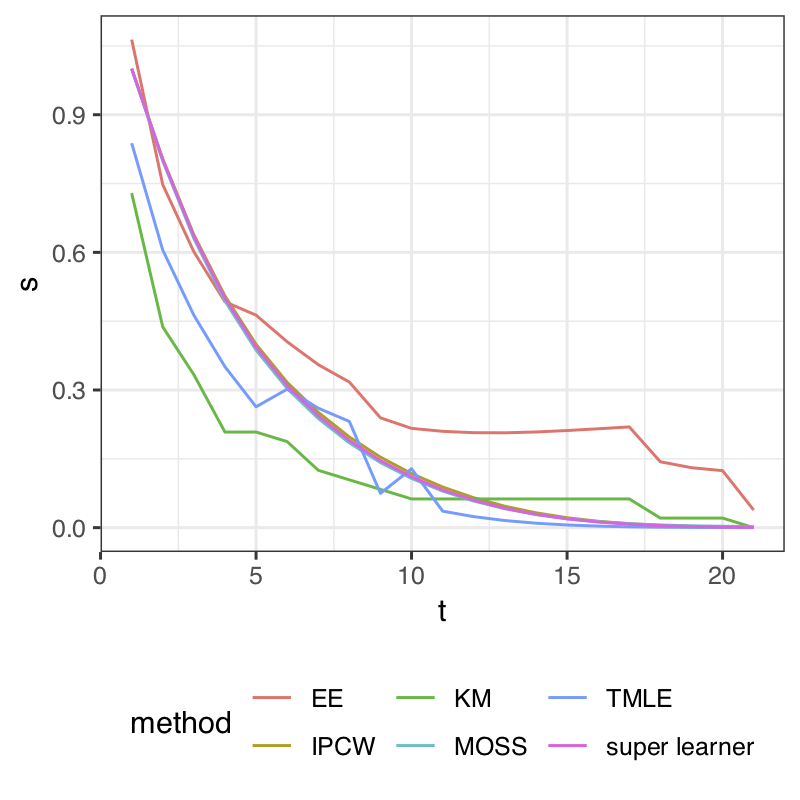}
(b)\includegraphics[width=.4\textwidth]{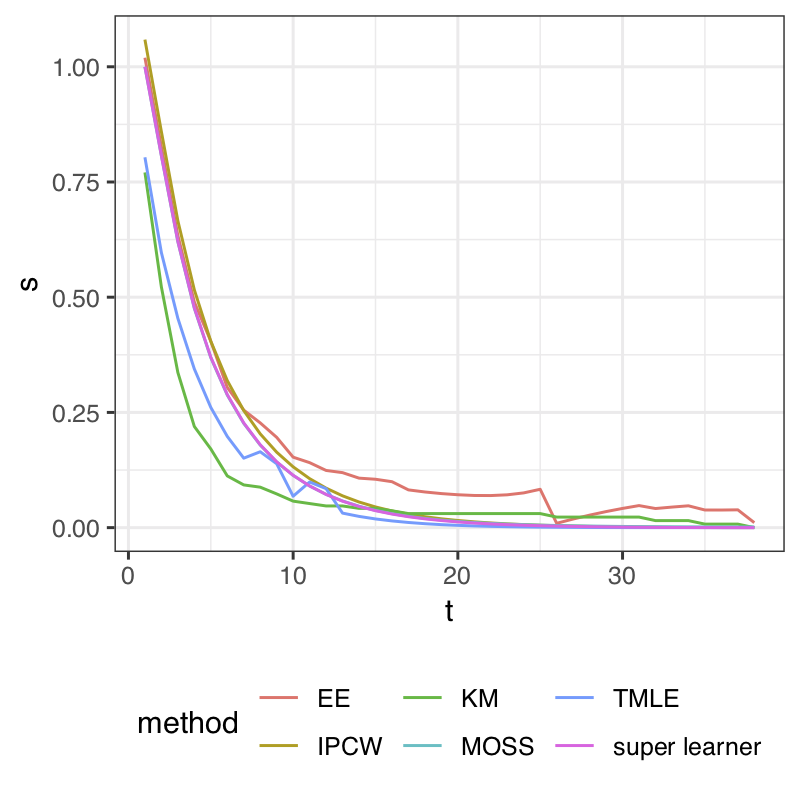}
\caption{Examples of non-monotone EE and TMLE estimators in simulation data of different sample sizes (plot a: n = 100, plot b: n = 1000). The target parameter is the marginal counter-factual survival curve for the treatment group ${\Psi _1}(P)$.}
\label{fig:once}
\end{figure}

The simulation results reflect what is expected based on theory. Figure \ref{fig:once} are examples in the simulation where the EE and classic TMLE methods do not produce monotone survival curves. 
Figure \ref{fig:sim} computes the metrics at different time points of the entire survival curve. One-step TMLE methods has lowest MSE under all sample sizes, with 33\% smaller MSE than the second best method (iterative TMLE) in small sample size. 
EE has a large variance in small sample size (n = 100) and its MSE becomes more comparable to iterative TMLE in larger sample size (n = 1000).
Kaplan-Meier is not consistent and has large MSE especially in large samples, although in finite samples its bias is not large compared to its variance.
IPCW has the largest variance and MSE under all sample sizes.
As sample size increases one-step TMLE converges to iterative-TMLE, and both TMLEs are better than IPCW, EE and Kaplan-Meier.

In Section \ref{sec:one_step} we gave intuition that the universal least favorable submodel can be viewed as a ridge logistic regression applied in the targeting step. Curious readers might be interested in the performance if we use a LASSO logistic regression instead. We also experiment this in the simulation (marked by `MOSS\_l1', while our proposed one-step TMLE is denoted `MOSS\_l2'), and we see that the difference between the two kinds of penalizations is small: both types of one-step TMLE outperforms iterative TMLE in finite sample and converge to iterative TMLE in the asymptotic. We find that using LASSO logistic regression improves MSE in large $t$ (where there are fewer samples) at a cost of a slightly larger MSE in small $t$. Therefore, we only recommend to use LASSO logistic regression for targeting step when minimax guaranteed improvement (across $t$) on the iterative TMLE is preferred.

\section{Data analysis} \label{sec:data_analysis}

To illustrate the finite sample performance of the one-step TMLE, we use a dataset from a classic monoclonal gammopathy study, an observational survival analysis dataset that first established the predictive relationship between the initial concentration of serum monoclonal protein and the progression to multiple myeloma or another plasma-cell cancer \citep{mgus}. For each subject, we define the (right-censored) outcome $\tilde T$ as the time until progression to a plasma cell malignancy or last contact, the treatment $A$ as the monoclonal spike on serum protein electrophoresis (1 = the spike is higher than 1.5 g/dL, 0 = the spike is lower than 1.5 g/dL), and include all baseline covariates $W$ (age, gender, hemoglobin, creatinine) that are measured upon enrollment of the subjects. The original study is on the predictive power of $A$ on the outcome and not the causal relationship, so there are definitely unmeasured confounders left out from this dataset. Nonetheless, we use the data to illustrate the statistical properties of different estimators. The trial measured 1338 complete cases after we discarded 46 subjects with missing data. We find that there is a practical violation of positivity assumption for time larger than 160 months. Therefore, we perform manual truncation of the dataset so that observations with follow-up time beyond 160 months are censored. We also transform the time unit of the dataset for ease of computation $\tilde T_{new} = \lceil \tilde T / 20 \rceil$, and we verify that this transformation does not change the scientific results. The preprocessed data contain 405 patients in the treatment group and 933 patients assigned to control.

\begin{figure}[ht]
    \centering
    \includegraphics[width=0.95\textwidth]{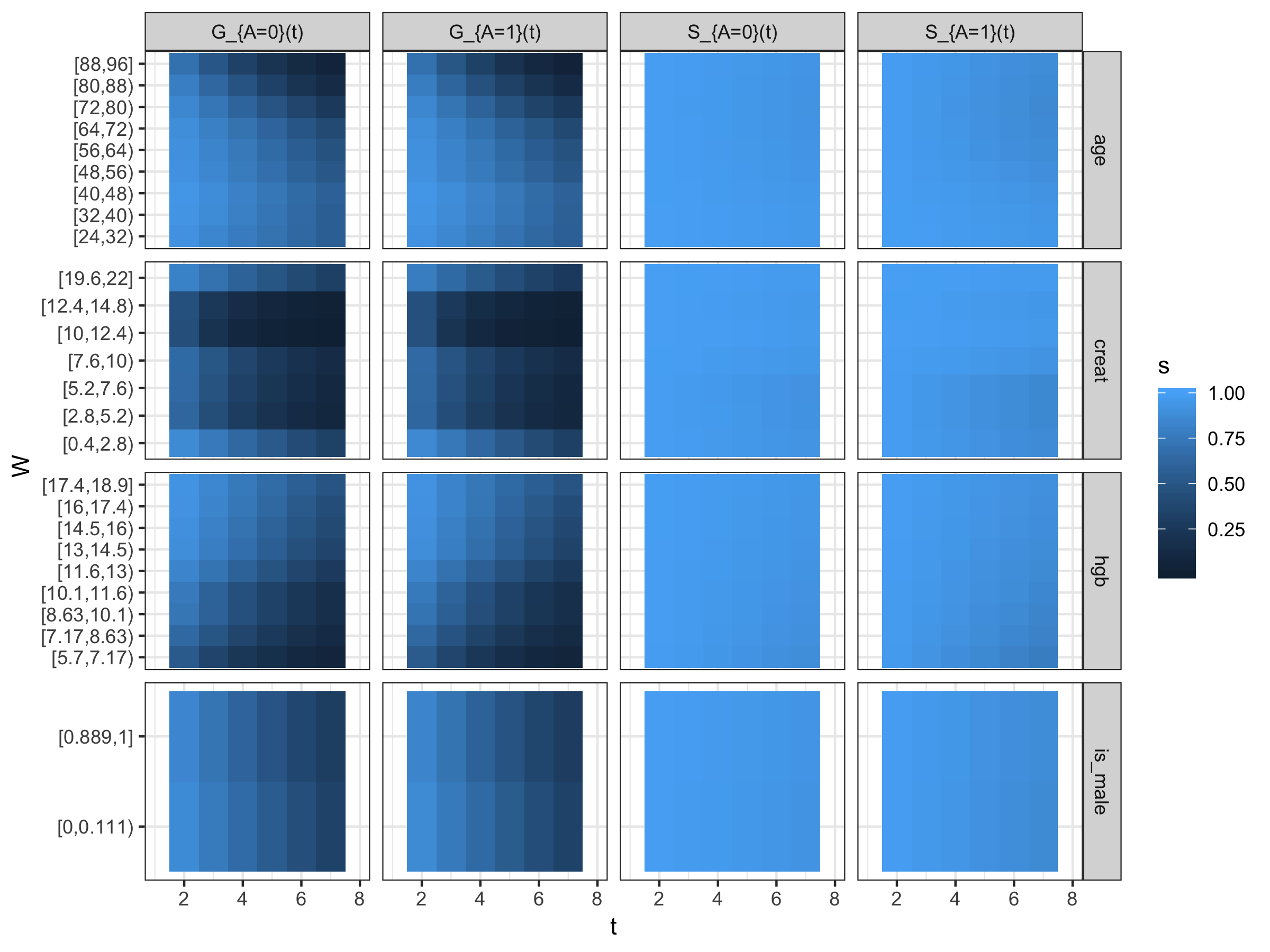}
    \caption{Partial dependency plots of the initial super learner fits for the conditional survival curves, where the y-axis is the baseline covariate value, the x-axis is time. Column 1 is the conditional survival of censoring event for control group; Column 2 is the conditional survival of censoring event for treatment group; Column 3 is the conditional survival of failure event for control group; Column 4 is the conditional survival of failure event for the treatment group. Row 1 plots have age on the y-axis; Row 2 plots have creatinine on the y-axis; Row 3 plots have Hemoglobin on the y-axis; Row 4 plots have gender indicator on the y-axis}
    \label{fig:sl}
\end{figure}

We first estimate the marginal survival curve for the treatment and control groups. We compare a plug-in parametric fit using GLM, plug-in SuperLearner fit, IPCW, EE, classic TMLE, and one-step TMLE targeting the whole curve. The SuperLearner initial fits combine main term generalized linear model, main term generalized additive model \citep{gam}, main term multivariate adaptive regression splines \citep{earth}, and random forest \citep{rf}. The same learner library is used for fitting the conditional survival for failure event and censoring event, as well as the propensity score.
The conditional survival functions estimated by SuperLearner \citep{superlearner} are presented in Figure \ref{fig:sl}. There is a complex interaction effect between baseline covariates (age and hemoglobin) and time in the conditional hazard of censoring event, so nonparametric methods such as SuperLearner is necessary.

\begin{table}[ht] \centering 
\begin{tabular}{@{\extracolsep{5pt}}lcccccc} 
\\[-1.8ex]\hline 
\hline \\[-1.8ex] 
Time & \multicolumn{1}{c}{Mean} & \multicolumn{1}{c}{St. Dev.} & \multicolumn{1}{c}{Min} & \multicolumn{1}{c}{Pctl(25)} & \multicolumn{1}{c}{Pctl(75)} & \multicolumn{1}{c}{Max} \\ 
\hline \\[-1.8ex] 
1 &  3.337 & 0.418 & 2.598 & 3.119 & 3.482 & 9.036 \\ 
2 &  4.049 & 1.395 & 3.040 & 3.560 & 4.078 & 27.047 \\ 
3 &  4.929 & 3.825 & 3.176 & 3.906 & 4.889 & 89.911 \\ 
4 &  6.592 & 11.662 & 3.361 & 4.372 & 6.248 & 319.789 \\ 
5 &  10.267 & 37.766 & 3.623 & 5.020 & 8.642 & 1,126.922 \\ 
6 &  19.899 & 129.833 & 4.012 & 5.920 & 12.971 & 4,095.776 \\ 
7 &  42.697 & 301.106 & 4.450 & 7.062 & 21.295 & 7,625.646 \\ 
8 &  85.334 & 422.724 & 4.941 & 8.526 & 37.284 & 7,625.646 \\ 
\hline \\[-1.8ex] 
\end{tabular} 
    \caption{Distribution of $\frac{1}{g(A=1 | W)G_C(t | A=1,W)}$ in the monoclonal gammopathy study} 
\end{table} 

\begin{table}[ht] \centering 
\begin{tabular}{@{\extracolsep{5pt}}lcccccc} 
\\[-1.8ex]\hline 
\hline \\[-1.8ex] 
Time &  \multicolumn{1}{c}{Mean} & \multicolumn{1}{c}{St. Dev.} & \multicolumn{1}{c}{Min} & \multicolumn{1}{c}{Pctl(25)} & \multicolumn{1}{c}{Pctl(75)} & \multicolumn{1}{c}{Max} \\ 
\hline \\[-1.8ex] 
1 & 1.437 & 0.054 & 1.124 & 1.403 & 1.472 & 1.626 \\ 
2 & 1.725 & 0.334 & 1.417 & 1.540 & 1.790 & 4.866 \\ 
3 & 2.062 & 0.847 & 1.489 & 1.675 & 2.126 & 14.574 \\ 
4 & 2.675 & 2.362 & 1.588 & 1.863 & 2.717 & 53.371 \\ 
5 & 3.949 & 7.395 & 1.720 & 2.135 & 3.698 & 195.817 \\ 
6 & 7.116 & 25.968 & 1.895 & 2.541 & 5.534 & 752.807 \\ 
7 & 14.750 & 66.451 & 2.091 & 3.027 & 8.788 & 1,417.807 \\ 
8 & 31.445 & 122.168 & 2.319 & 3.624 & 15.079 & 1,504.796 \\ 
\hline \\[-1.8ex] 
\end{tabular} 
  \caption{Distribution of $\frac{1}{g(A=0 | W)G_C(t | A=0,W)}$ in the monoclonal gammopathy study} 
\end{table} 

\begin{figure}[ht]
    \centering
    \includegraphics[width=0.95\textwidth]{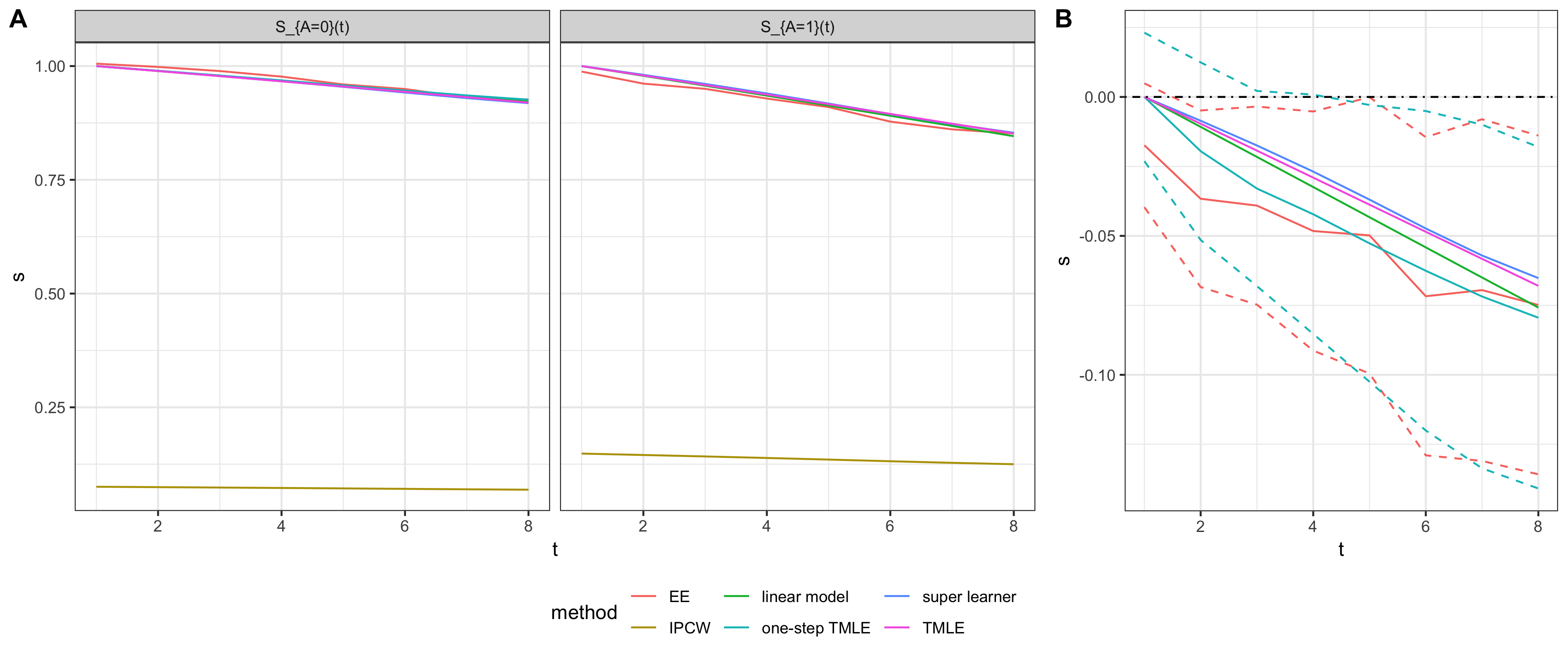}
    \caption{Results for different counterfactual survival curve estimators on the Monoclonal gammopathy data. Panel A is survival curve estimates for the control group and treatment group, using different estimators. Panel B is the difference curve in survival probabilities (treatment group minus control group), using different estimators.}
    \label{fig:realdata_panel2}
\end{figure}

Figure \ref{fig:realdata_panel2}(A) shows the different estimators' results for the treatment and control group survival curves. The one-step TMLE, TMLE, EE and SuperLearner fits are close to each other, suggesting that the dataset is large. EE is slightly not monotone for the treatment group survival curve. IPCW is drastically different from all other estimators, which is the worst performing method.
Second, the delta method is applied to obtain the estimators for the difference in survival probabilities (treatment minus control). Wald 95\% confidence bands for EE and TMLE are calculated using the efficient influence curve. SuperLearner is different from the parametric fit, suggesting that nonparametric regression is crucial for this analysis. EE is not monotone.
Lastly, to check how well the estimators perform in a finite sample, we randomly subsample the pre-processed data into smaller sizes and re-compute all methods. The procedure is repeated 100 times, and we count how frequent each estimator yields a monotone curve. The percentages are reported in Table \ref{tbl_monotone1} and \ref{tbl_monotone0}. We find that EE has the highest probability of becoming not monotone when all other conditions held equal. Classic TMLE outputs a monotone survival curve at least 80\% of the times, and one-step TMLE is guaranteed to be monotone.

\begin{table}[ht]
\begin{minipage}{0.45\textwidth}
\begin{tabular}{rrrr}
\hline
n    & EE    & TMLE  & one-step TMLE \\
\hline
100  & 42\%  & 91\%  & 100\%         \\
500  & 74\%  & 93\%  & 100\%         \\
1000 & 100\% & 100\% & 100\%        \\
\hline
\end{tabular}
\caption{For each method and subsample size, the percentage of experiments when the estimator outputs a monotone survival curve in the monoclonal gammopathy study (for the treatment group).}
\label{tbl_monotone1}
\end{minipage}
\hfill
\begin{minipage}{0.45\textwidth}
\begin{tabular}{rrrr}
\hline
n    & EE    & TMLE  & one-step TMLE \\
\hline
100  & 38\%  & 81\%  & 100\%         \\
500  & 90\%  & 93\%  & 100\%         \\
1000 & 100\% & 100\% & 100\%        \\
\hline
\end{tabular}
\caption{For each method and subsample size, the percentage of experiments when the estimator outputs a monotone survival curve in the monoclonal gammopathy study (for the control group).}
\label{tbl_monotone0}
\end{minipage}
\end{table}

\section{Discussion} \label{sec:discussion}

In this paper, we provided a one-step TMLE for estimating the treatment-rule specific survival curve while targeting the entire survival curve at once. The one-step estimator has implications for the survival analysis literature by allowing one to construct a TMLE for the infinite dimensional survival curve in a single step. The new method is asymptotically linear and efficient, just as the iterative TMLE, which adjusts for baseline covariates and accounts for informative censoring through inverse weighting. Additionally, the one-step estimator targeting the entire survival curve respects the monotonically decreasing shape of the estimand. On top of that, the new TMLE for the entire curve also yields a fully compatible TMLE for any function of the whole survival curve, such as the median, quantile, or truncated mean. Thus there is no need to compute a new TMLE for each specific feature of the survival curve, or difference of survival curves. All of these advantages come without requiring any parametric modeling assumptions and is robust to misspecification of the hazard fit. Our simulation confirms the theory in existing literature: that in situations where targeting is difficult due to extreme propensity scores, using one-step TMLE that fluctuates universal least favorable submodel may provide robustness and efficiency over iterative TMLE. Under large sample sizes, iterative and one-step TMLE are comparable. We show that in practical finite sample situations for survival analysis, using universal least favorable submodel to target a multi-dimensional or even infinite-dimensional target parameter is likely to result in a more efficient and stable estimator. It is not clear how our methods compare with applying isotonic regression to the curve defined by the one-step TMLEs targeting one survival probability across all time-points. This represents another valid and possible method to consider if getting the whole survival curve is the goal of the analysis.

\bibliographystyle{plainnat}
\bibliography{merged}


\section*{Supplementary Materials} 

The R software that implements the methodologies and reproduces the analyses in this paper, is available with this paper at the Biometrics website on Wiley Online Library and open-sourced online \citep{MOSS,MOSS_simu}.

\end{document}